\documentclass[12pt]{iopart}
\usepackage{iopams}

\expandafter\let\csname equation*\endcsname\relax
\expandafter\let\csname endequation*\endcsname\relax 

\usepackage{subfig}
\usepackage{amsmath}
\usepackage{multicol}
\usepackage{subfig}
\usepackage[normalem]{ulem} 
\usepackage{cite}
\newcommand{\cch}[1]{\left[#1\right]}
\newcommand{\cha}[1]{\left\{#1\right\}}
\newcommand{\prt}[1]{\left(#1\right)}

\usepackage[colorinlistoftodos]{todonotes}
\usepackage[hidelinks,colorlinks,breaklinks=true]{hyperref}
\hypersetup{
  colorlinks=true, linkcolor=blue, 
  citecolor=blue,   
  filecolor=magenta,
  urlcolor=blue,
  bookmarksdepth=4
}

\begin{document}

\title{Transport threshold in a quantum model for the KscA ion channel }

\author{N. De March, S. D. Prado and L. G. Brunnet} 

\address{Instituto de F\'{\i}sica, Universidade Federal do Rio Grande do Sul (UFRGS) \\
CP: 15051 Porto Alegre, RS, Brazil}
\ead{nicmarck@gmail.com}
\vspace{10pt}
\begin{indented}
\item[]\today
\end{indented}

\begin{abstract}
The mechanism behind the high throughput rate in K$^{+}$ channels is still an open problem. Recent simulations have shown that the passage of potassium through the K$^{+}$ channel core, the so-called selectivity filter (SF), is water-free against models where the strength of Coulomb repulsion freezes ions conduction. 
It has been suggested that quantum coherent hopping might be relevant in mediating ion conduction. Within the quantum approach and the hypothesis of desolvated ions along the pathway, we start with a number of particles in a source to see how they go across the SF modeled by a linear chain of sites to be collected in a drain.
As a main result we show
that there is a threshold  SF occupancy is three ions on average, which is in agreement with recent classical model simulations.

\end{abstract}


\vspace{2pc}

\submitto{\JPCM}

\maketitle

%
\section{Introduction}

\begin{figure}[h!]
\centering
\subfloat[]{{\includegraphics[width=7.2cm,clip=true]{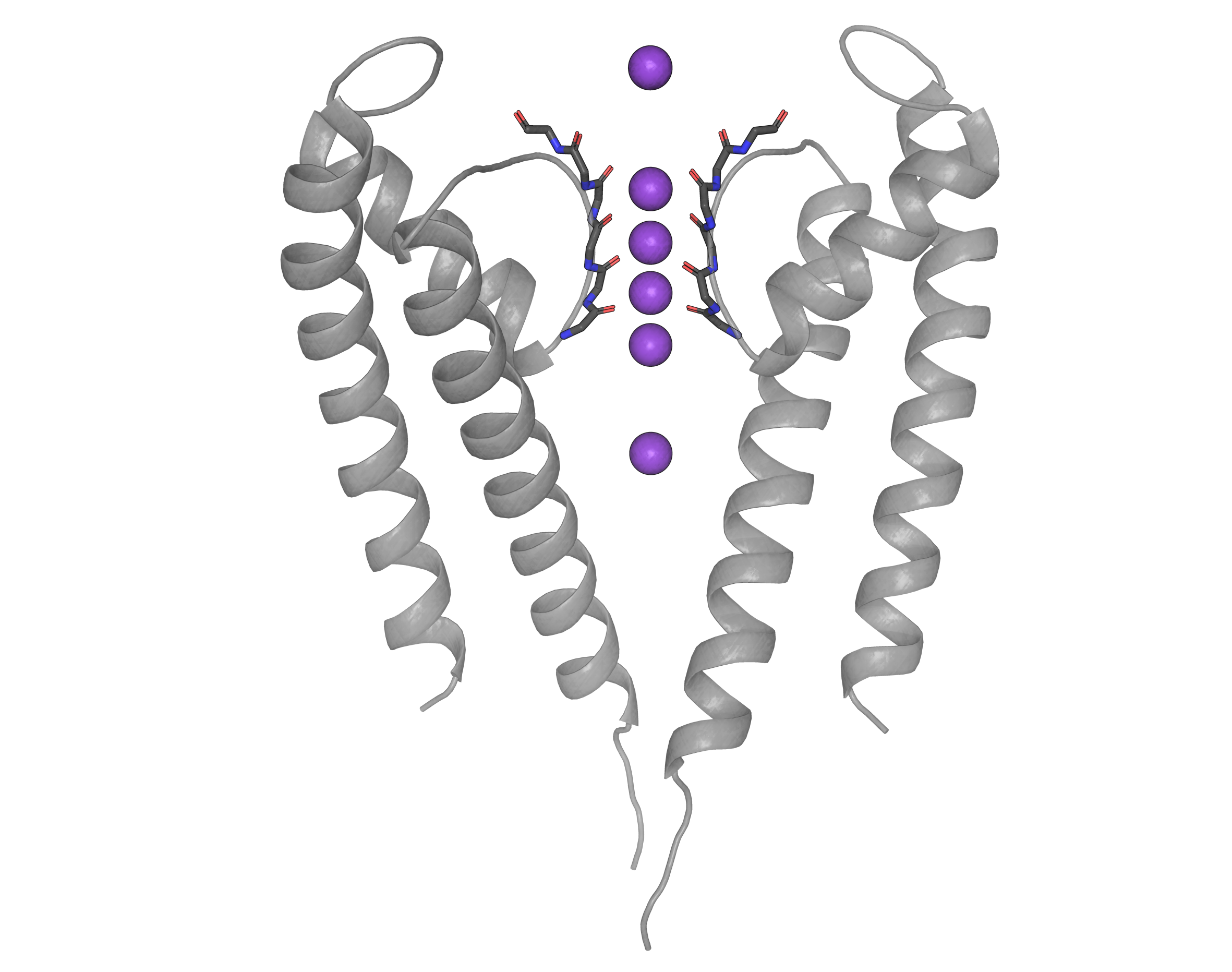} }}%
\qquad
\subfloat[]{{\includegraphics[width=7cm,clip=true]{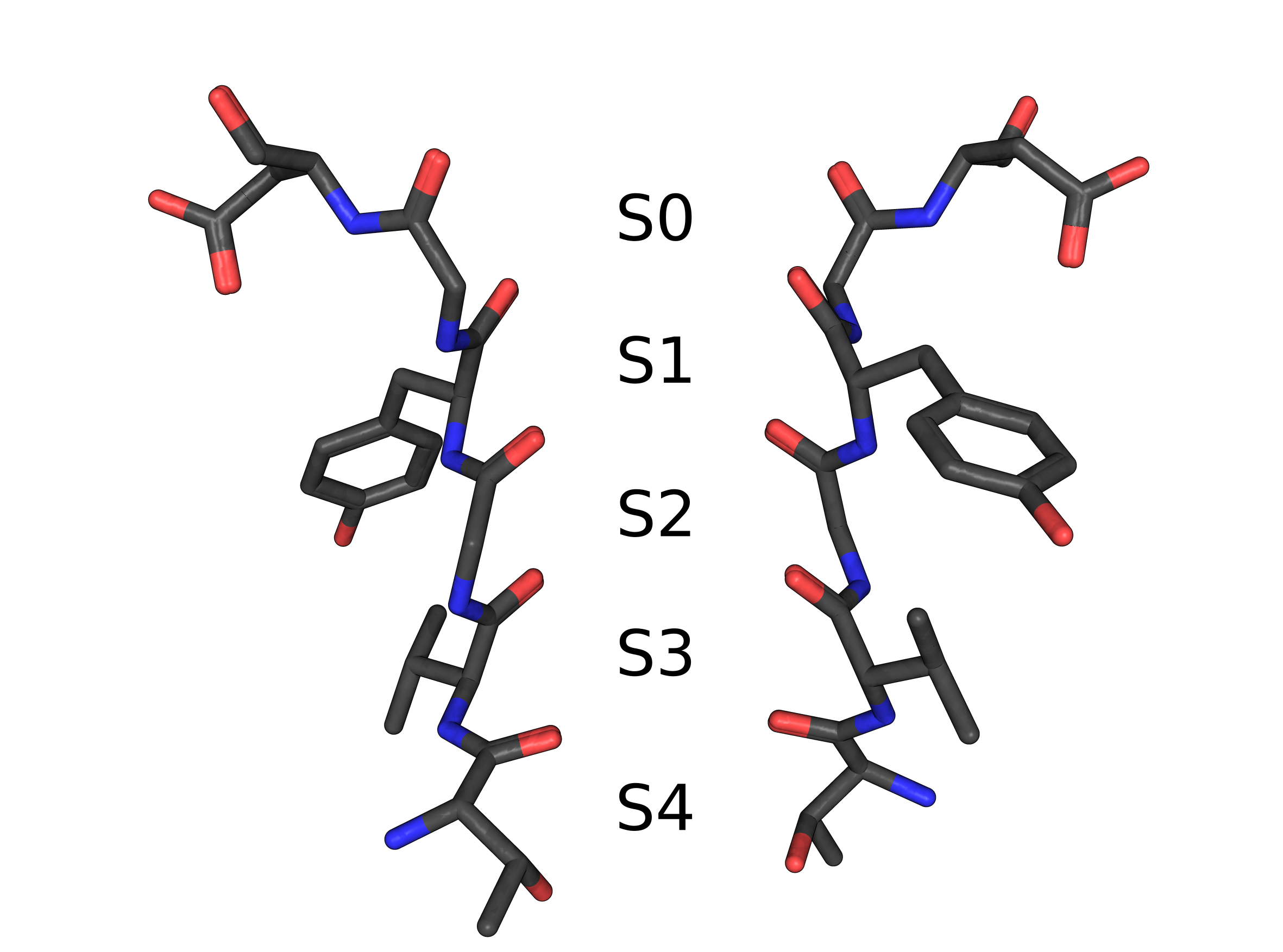} }}%
\caption{(a) Crystal structure of KcsA channel. (b) Selectivity filter binding sites (S1 to S4) composed of carbonyl groups, oxygen atoms (in red) point towards the ion passage}
\label{kcsachannel}
\end{figure}

Potassium ion channels strongly discriminate K$^{+}$ ions, allowing their passage across cell membrane at a near-diffusion-limited conduction efficiency ($10^8$ ions/s). Doyle et al. \cite{doyle1998structure} and Zhou et al. \cite{Zhou2001} determined the crystal structure of the bacterial potassium ion channel KcsA ({\it Streptomyces lividans}) Fig. \ref{kcsachannel}(a), shedding light on the chemistry  of the selectivity filter. A central characteristic is the highly conserved signature sequence of amino acids (TVGYG). 

The SF is the narrowest part of the conduction pathway, and it is responsible for high selectivity and fast permeation. Ions leave the water cavity and are desolvated when entering the filter. Due to the SF spatial constriction, ions align in a single-file fashion. The carbonyl oxygen atoms of the SF backbone interact with the ions via four consecutive K$^{+}$ binding sites (Figure \ref{kcsachannel}).

Several studies investigate K$^{+}$ channels using different techniques, such as electrophysiology \cite{doyle1998structure,Zhou2001,thompson2009mechanism,ranjan2019}, spectroscopic methods \cite{kratochvil2016instantaneous}, x-ray crystallography \cite{doyle1998structure,Zhou2001}, and molecular dynamics simulations \cite{kopfer2014ion,kopec2018direct}. However, even with all these accomplishments, the mechanism under ion permeation and the achieved high potassium ion throughput rate remains unclear. In the past, it was assumed that the K$^{+}$ ions co-migrated with water molecules through the selectivity filter, with the conduction configuration consisting of K$^{+}$ ions and water molecules in alternation, one after the other \cite{Zhou2001,zhou2003occupancy,thompson2009mechanism}. This hypothesis lies on the idea that the Coulomb repulsion between two nearby desolvated ions would freeze conduction. Surprisingly,  recent findings contradict this water-mediated conduction mechanism \cite{kopfer2014ion,kopec2018direct}. Molecular dynamics simulations show that the primary mechanism for ion conduction is the direct contact between neighboring potassium ions (direct knock-on). Furthermore, experimental findings \cite{Oster2019} revealed no water in the selectivity filter under physiological conditions. 

Motivated by the recent achievements in Quantum Biology \cite{Marais2018}, it was suggested that quantum effects might be relevant in mediating ion conduction and selectivity in the SF.
Vaziri and Plenio \cite{vaziri2010quantum} argued that ion transport and selectivity might involve quantum effects.
Besides that, Vaziri and Plenio \cite{vaziri2010quantum} and Ganim et al. \cite{ganim2011vibrational} proposed that  quantum coherence in the K$^{+}$ ion channels might be due to the SF backbone flexibility, so they suggest investigating its vibrational excitations. Based on that, some authors analyzed how noise and disorder affect the ion transport and selectivity through the excitation energy transfer in the structure \cite{toymodel, jalalinejad2018excitation}. Also using noise, Cifuentes and Semi\~ao \cite{cifuentes2014quantum} argued that  quantum interference between two possible pathways might be advantageous. Searching for the possibility of quantum superposition, Salari et al. \cite{salari2015quantum} investigated decoherence times for filter states  using  data obtained via MD simulations. They concluded the dynamics is long enough coherent for interference to leave traces on the SF conformation.

Furthermore, Salari et al. \cite{salari2015classical}
claim that there is no statistical correlation between the classical motion of carbonyl groups and the high ionic currents. In \cite{salari2017quantum}, Salari et al. compared two neighboring ion channels to a double-slit experiment and examined the  interference of potassium ions going through both channels. They argued that this quantum interference could explain potassium ion selectivity.

Summhammer et al. \cite{summhammer2012quantum} described a single potassium ion motion and its interaction with the surrounding carbonyls dipoles in the SF  analyzing the Schr\"odinger equation solutions. More recently, in \cite{summhammer2018quantum,summhammer2020quantum} Summhammer and coworkers simulated the transmission of K ions using classical molecular dynamics (MD) at the atomic scale together with a quantum mechanical version of MD simulation. Besides all  that, there are results arguing  about possible effects of quantum coherence at the onset dynamics of propagating voltage pulses (action potentials) in neurons  \cite{bernroider2012,moradi2015study,quantum1020026}.

Assuming a quantum perspective, we will extend our previous two-particle model, wherein we focused on the Coulomb repulsion between ions in their conduction through the SF. We showed that Coulomb strength dominates the process timescale \cite{de2018coulomb}.
This assumption agrees with the recent findings that there is no water involved in the potassium ion permeation. In addition, observations obtained via crystallography and molecular dynamics simulations showed configurations having at least two particles inside the filter, with no more than three \cite{kopfer2014ion}. Thus, in this work, we explore increasing the number of particles  in order to produce a higher filter population on average. Also, we observe the implications in the system characteristic timescales.

\section{Model \& Methods}\label{section2}

\subsection{Model for SF structure and transport}

The selectivity filter comprises four potassium binding sites (S1-S4), as well as a fifth K$^{+}$ binding site (S0) observed at the external side \cite{berneche2001energetics}, as shown in Fig. \ref{kcsachannel}(b). 
A distance of about 0.24 nm separates two neighbors potential minima, so that each ion has to go across a barrier between two SF trapping sites. The barrier height depends both on the presence of an ion at a specific site and on the protein thermal vibration. According to molecular dynamics simulation, this  height fluctuates from  1.7 to 8.0 $k_{B}T$, approximately \cite{gwan2007cooperative}. Given the ion thermal wavelength and the SF characteristic dimension it has been suggested that ion transport could be explained by quantum tunneling through a potential barrier between neighboring sites, and that the quantum coherence might be significant \cite{vaziri2010quantum}. Besides other features, the authors have analyzed the efficiency for a single particle model in a chain of five sites with periodic boundary conditions. We extend their model starting with more particles in a source aiming to study the effect of Coulomb repulsion when particles go through the channel. We evaluate the rates of ion transmission using the selectivity filter data reported in the literature \cite{vaziri2010quantum, berneche2001energetics, kopfer2014ion, gwan2007cooperative}.

Assuming the  simplest approximation of a rectangular  barrier of width $\Delta$, the tunneling probability is $p_{tun}\sim e^{-\Delta\sqrt{2m\Delta E/h^{2}}}$, where $\Delta E$ is the difference between the barrier energy and the total ion energy and $m$ is the potassium mass. The ion energy is its thermal energy, the barrier height is assumed to be the lowest value in the estimated range, $1.7k_{B}T$, and the barrier width $\Delta$ has to be smaller than $0.24$ nm, the distance between successive sites.
The ion kinetic energy, $K$, can vary from 0.5 to 1.7 $k_BT$  \cite{kopfer2014ion}, so that we can estimate the range of the trapping frequency,  $\nu=K/h$. Therefore, the estimated tunneling rate, $\nu~ p_{tun}$, can vary from $10^{11}$ to $10^{13}\,\text{s}^{-1}$.\footnote{Note that the terms tunnelling and hopping are used as synonymous in this paper.  The potential axial separation of the minima and the barrier height fluctuate based on the presence of an ion at a particular site. The transmission involves quantum tunnelling through the potential barrier between individual neighbouring binding sites. In this simplified model, one estimates the tunnelling probability, and from that the hopping rates to be used in a tight-binding type Hamiltonian \cite{vaziri2010quantum}.}
 
 We approximate the SF as a linear chain of N=5 binding sites labeled from 1 to 5 towards the  flux direction. The intracellular site (S4 in Fig. \ref{kcsachannel}) has index-1 and the extracellular site has index-5 (S0 in Fig \ref{kcsachannel}). We use a tight-binding Hamiltonian with an inter-site Coulomb term to model the system,

\begin{equation}
  H= - \hbar c
 \sum_{j=1}^{N-1}  \prt{{\sigma}_{j}^{\dag}{\sigma}_{j+1}+{\sigma}_{j+1}^{\dag}{\sigma}_{j}} +\frac{1}{2}\sum_{j=1}^{N}\sum_{j'\neq j}^{N}\frac{U}{|j-j'|}n_{j}n_{j'},
\label{tight}
\end{equation} where $\hbar$ is the Planck's constant, $c$ is the hopping rate between adjacent sites, and $j$ is the  site label.  ${\sigma}_{j}^{\dag}$ (${\sigma}_{j}$) is the creation (annihilation) operator at site $j$. U is the Coulomb strength and $n_{j}$ is the number operator $\sigma_{j}^{\dag}\sigma_{j}$. Given that $ \hbar c\ll U \approx 5.14\,\text{eV}$, we can define an effective hopping, $c_{eff} \approx  \hbar c^{2}/U$, obtained from first-order perturbation theory \cite{de2018coulomb}.
Once the system time scale is proportional to $c_{eff}$, the higher the Coulomb repulsion, the slower the dynamics.

In our simulations,  we will have  a source supplying particles that is the first SF site and a drain  connecting the last chain site, where the ions are collected. So, cytoplasm potassium ions enter the SF through site-$0$ and leave at site-$6$. Lindblad operators are appropriate tools for this kind of model \cite{de2018coulomb,breuer2002theory}:

\begin{equation}
\mathcal{L}_{s}\prt{\rho}=\Gamma_{s}\prt{-\cha{{\sigma}_{0}^{\dag}{\sigma}_{1}{\sigma}_{1}^{\dag}{\sigma}_{0},\rho}+2{\sigma}_{1}^{\dag}{\sigma}_{0}\rho{\sigma}_{0}^{\dag}{\sigma}_{1}},
\label{lsource}
 \end{equation}where $s$ stands for source,  $\Gamma_{s}$ denotes source supplying rate and $\rho=\rho(t)$ is the density matrix. Analogously to Eq. (\ref{lsource}), we use for the drain the Lindblad operator,
\begin{equation}
\mathcal{L}_{d}\prt{\rho}=\Gamma_{d}\Biggl(-\cha{{\sigma}_{N}^{\dag}{\sigma}_{N+1}{\sigma}_{N+1}^{\dag}{\sigma}_{N},\rho}+2{{\sigma}_{N+1}^{\dag}{\sigma}_{N}  \rho{\sigma}_{N}^{\dag}{\sigma}_{N+1}}\Biggr),
\label{ldrain}
\end{equation}
where $d$ stands for drain and  $\Gamma_{d}$ denotes the leaking rate of the drain.

In order to do the computation, we use fermion number occupation basis for the ion sites that run from $N$=1 to 5. This basis is very conveninent to simulate the situation where there is no two particles at the same site. On the other hand, source and drain are treated on a boson occupation basis since   they can be arbitrarily populated. We obtain the system evolution by integrating the Lindblad master equation \cite{breuer2002theory,rivas2012open} 
\begin{equation}
 \frac{\mathrm{d}}{\mathrm{d} t}{\rho}\prt{t}= -\frac{i}{\hbar}\cch{H,\rho\prt{t}}+\mathcal{L}_{s}\prt{\rho}+\mathcal{L}_{d}\prt{\rho},
\label{lidbladdiago2}
\end{equation}which defines a set of  coupled linear ordinary differential equations. We perform numerical integration of the master equation with Lindblad terms (Eq. \ref{lidbladdiago2}) using the LAPACK (Linear Algebra Package)  library \cite{anderson1999lapack}.

\section{Numerical Results}

All results presented in what follows are rescaled by  $c_{eff}$, its inverse defining a typical timescale. We set Planck constant $\hbar=1$ and hopping rate $c=1$ for simplification.
Then time is $\tau=t\,c_{eff}$,  the rates are $\tilde{\Gamma}_s =\Gamma_s/c_{eff}$ and $\tilde{\Gamma}_d =\Gamma_d/c_{eff}$. The energy scale is $\hbar c $ and  $~{\cal U} = U /\hbar c~$ is in the range $[10^3 - 10^5]$ ~ \cite{de2018coulomb}. 
We verified that as in the two-particle system \cite{de2018coulomb}, the rates
  $\tilde\Gamma_s=\tilde\Gamma_d=1$ also minimizes the Coulomb repulsion for a system with multiple particles. Consequently, we kept these rates in all the simulations. Finally, for the initial state the source is fully occupied by $N_{tot}$ particles.

In section \ref{section2} we argue  that  $ U \gg \hbar c$ leads to an effective hopping, $c_{eff}\approx \hbar c^2/U$, that defines the system characteristic time. In Ref. \cite{de2018coulomb}, the results for a two-particle system was in agreement with this approximation showing that the Coulomb repulsion dictates the system timescale. We show here similar behavior for a larger number of particles.

\begin{figure}
\centering
\includegraphics[width=1.0\linewidth,clip=true]{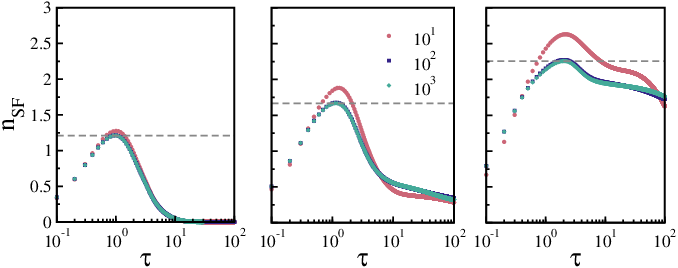} 
\caption{(Color online) Selectivity filter occupancy versus reduced time (logarithmic scale in $\tau$-axis) for different values of Coulomb repulsion $U$, and $\tilde{\Gamma}_{s}=\tilde{\Gamma}_d=1$. Two-particle system (left), three-particle system (middle), and five-particle system (right). The curves matching shows a good time rescaling with Coulomb repulsion for different numbers of particles in the system. Also, the maximum SF occupancy for each total particle number decreases for increasing Coulomb repulsion, saturating in a minimum at $U=10^2$ (gray dashed line). }
\label{235numberparticlesdifUblind}
\end{figure}

Fig. \ref{235numberparticlesdifUblind} (left) shows the selectivity filter occupancy $n_{SF}$ for a two-particle system versus reduced time. Curves represent different $U$ values. This system could be the most simple case to analyze the effects of Coulomb repulsion on the ion transmission, but since the parameters are chosen for optimal efficiency,  the occupancy  never reaches, on average, two particles inside the filter so that the particles interaction is less effective than initially supposed. So, focusing on the role played by Coulomb repulsion, we added particles to the source. Now, with a larger mean SF population, we can compare our simulations with the recent literature that reports results showing an SF occupancy up to three potassium ions for more than 1300 spontaneous K$^{+}$ permeation events for approximately 50  $\mu$s under varying  K$^{+}$ concentrations \cite{kopfer2014ion,Oster2019,Langan2020}. 

 We vary the Coulomb strength $U$ in the range of interest  starting with two, three, and five particles in the source. Fig. \ref{235numberparticlesdifUblind} (left, middle and right respectively) depicts SF occupancy versus reduced time for 3 values of $U$ as indicated in the legend. As the number of particles increases,  Fig. \ref{235numberparticlesdifUblind} (middle and right), the overlap  among the curves gets better as $U$ gets larger. This result shows that for $U\gtrsim 100$, the Coulomb repulsion defines the typical time scales  also for sources with more than 2 particles \cite{de2018coulomb}. Besides that, the SF occupation  reaches a limiting value as the repulsion strength increases, as depicted in the gray dashed line.
 
The time rescaling with $U$ allows a generalization to systems with larger sources  ($N_ {tot}$). Therefore, it is possible to infer the properties of the  larger system with large  $U$  by simulating one with a lower $U$ that requires less  computational time. Fig. \ref{235numberparticlesdifUblind} shows that for a fixed  $N_{tot}$ particles system, we can infer the timescale of 
$U=10^3$ from $U=10$, and the SF occupancy $n_{SF}$ of $U=10^3$ from $U=10^2$.

\begin{figure}
\centering
\includegraphics[width=0.6\linewidth,clip=true]{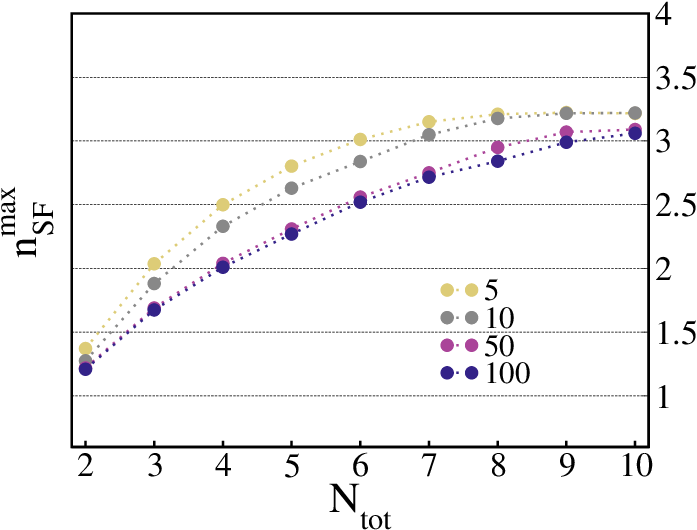}
\caption{ (Color online)  Maximum SF occupancy, $n_{SF}^{max}$, as function of the number of particles, $N_{tot}$, for different values of $U$ (shown in the legend) and $\tilde{\Gamma}_{s}=\tilde{\Gamma}_{d}=1$. The maximum SF occupancy, $n_{SF}^{max}$ tends to a number close to 3 for higher U and asymptotic large sources in agreement with recent water-free classical model simulations \cite{kopec2018direct}.
}
\label{maxinsideu10}
\end{figure}

Fig. \ref{maxinsideu10} shows the maximum SF occupancy, $n_{SF}^{max}$, as a function of the number of particles, $N_{tot}$, for different $U$ values. Note that, as in Fig. \ref{235numberparticlesdifUblind},  $n_{SF}^{max}$ decreases with $U$ to a limiting value.  The SF occupancy increases as $N_{tot}$ increases  but it tends to saturate at about 3 particles at a time.
To reach the maximum SF occupancy found in water-free permeation, we look for an $n_{SF}^{max}$ value for a system with high $N_{tot}$ values and  with $U\sim 10^3$. Since $n_{SF}^{max}$ saturates with increasing $N_{tot}$ and $U$, we can infer from Fig. \ref{maxinsideu10} that when $N_{tot}\geq 9$ and $U=100$, $n_{SF}^{max}$ is close to three, which is in agreement with recent water-free classical model simulations \cite{kopec2018direct}.


\begin{figure}
\centering
\includegraphics[width=0.6\linewidth,clip=true]{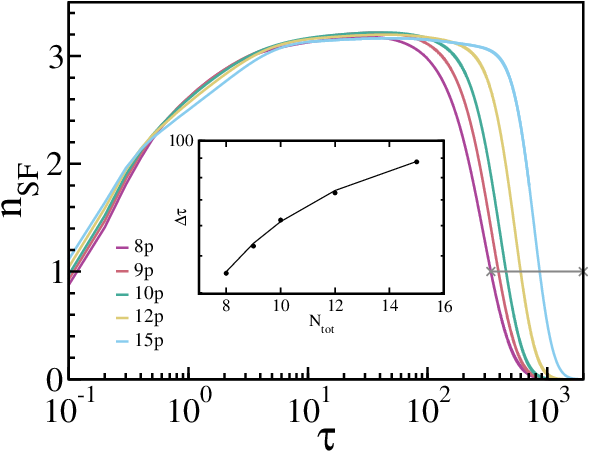}
\caption{(Color online) Selectivity filter population  versus reduced time for $N_{tot}\geq 8$ particles and $U=10$.
The time that SF stays in an almost steady-state population increases as we add particles to the system. The time increment in the plateau lag, $\Delta\tau$ is found measuring the time $\tau$ at which the SF remains with one particle $n_{SF}(\tau)=1$ (grey horizontal line). The inset shows  the $\Delta\tau$ (logarithmic scale in the $\Delta\tau$-axis) for $N_{tot}$ (black dots). A nonlinear regression fit for these data (continuous line) shows that for $N_{tot}$ large and $U=10^3$, we find one particle leaving the SF in $\sim  10^{- 8}\,s$, in agreement with the measured characteristic transport  times  for potassium ions \cite{doyle1998structure}. 
}
\label{numberparticlesu10blind}
\end{figure}

Figure \ref{numberparticlesu10blind} shows the selectivity filter occupancy $n_{SF}$ versus reduced time on a logarithmic scale. Curves represent different particle numbers in the system, $N_{tot}$ (indicated in the legend) for  $N_{tot}\geq 8$. We set $U=10$, since we can infer from this Coulomb strength the same qualitative results for $U=10^3$, as previously discussed. Note that, for $N_{tot}\geq 8$, $n_{SF}$ quickly saturates at its maximum value with the SF remaining filled in a nearly constant population, corroborating Fig. \ref{maxinsideu10}. 
 In order to estimate the time extension of the plateau lag as the number of particles increases, we count the time difference at the plateau elongation as we add one more particle to the system. For that, we record the time when a single particle remains in the SF, that is the time $\tau(N_{tot})$ when $n_{SF}=1$. This time is indicated by the intersection of the horizontal grey line with the occupation lines in Fig. \ref{numberparticlesu10blind}. We use this recorded time to evaluate $\Delta\tau$ that is, the lag time as we add a particle in $N_{tot}$. So, $\Delta\tau(N_{tot})=\tau(N_{tot})-\tau(N_{tot}-1)$, when $n_{SF}=1$. The inset of Fig. \ref{numberparticlesu10blind} shows the lag time increase, $\Delta\tau$, plotted against $N_{tot}$ (black dots). The continuous line is the nonlinear regression fit for these data.
 To estimate the maximum value of $\Delta\tau(N_{tot})$ we use a nonlinear regression $\Delta\tau=a(1-e^{-N_{tot}/c})+b$, with $a=195.57$, $b=-74.23$ and $c=8.5$ (inset in Fig. \ref{numberparticlesu10blind}, continuous line for the indicated $N_{tot}$). The maximum $\Delta\tau\sim 121$, arising when $N_{tot}\rightarrow\infty$. Rescaling $\Delta\tau$ for $U=10^3$, given that $t\equiv {\cal U} \tau/c$ we obtain $ \Delta t \approx 121\times 10^{3} \times 10^{-13} \sim \times 10^{-8}\,s$ for the lag time.
   This lag time obtained in the limit of large sources can be interpreted as the ion
typical crossing time, and  $\Delta t ~10^{-8}$s is a time corroborated by experiments \cite{gouaux2005principles}.


\section{Conclusions}

Recent experimental findings  show that under physiological conditions the primary mechanism for ion conduction is the direct contact between potassium ions \cite{Oster2019}, opposing the well-accepted water-mediated knock-on mechanism proposed based on crystallographic data \cite{morais2001energetic}. In a quantum model with desolvated ions we show that the Coulomb repulsion  plays the main role in setting the transport timescale slowing down the dynamics. However, this decrease can be mitigated by equating the input/output rates  to the effective hopping rate, thereby optimizing ion transport. 

In consonance with the recent literature \cite{kopfer2014ion,Oster2019,Langan2020} that shows a filter occupancy of two (up to three) ions at a time, we found that, as the particle number in the system increases, the filter population reaches a maximum close to three in the regime of strong Coulomb repulsion as it is the case for the SF. This being the main result of this paper. 
Moreover, using the relations imposed on the time scale by the Coulomb repulsion, we found the typical particle crossing time to be $\sim  10^{- 8}\,s$ for asymptotic large sources. So, adopting  an independent approach, we found the same results of SF population and ion throughput time as the ones from classical simulations \cite{kopec2018direct}.

It is worth emphasizing that to improve  this  model  we have to take into account  vibration mode effects in the carbonyl groups and/or thermal noises. They are important ingredients since they do imply a  shorter decoherence time limiting the time the dynamics is coherent \cite{vaziri2010quantum, ganim2011vibrational}. Accordingly, we aim to consider the oscillation of the carbonyl groups in future work.

\ack 
We thank the Brazilian agencies CAPES, CNPq, and
FAPERGS for financial support. We also acknowledge the Computational Center of the Physics Institute of UFRGS for cluster computing hours.

\clearpage

\bibliography{paper_main}               
\end{document}